# The study of water shell structure of double-helical B-DNA fragments using parallel computing.

A.V. Teplukhin, Yu.S. Lemesheva

# Introduction

Depending on the nucleotide sequence and environmental conditions local structure of the DNA double helix can vary within wide limits. In turn, particular secondary structure modulate local flexibility and curvature of the DNA duplex axis that are necessary for the biochemical processes. In particular, the domains of the poly (dA): poly (dT) type (so-called A-tract) has over 15 years are at the center of attention of scientists. It is known that duplexes containing A-tract, have many unusual properties compared to fragments of the native B-DNA. It is believed that this is due to the ability of A-tracts to adopt stable B'-conformation, characterized by a narrowed glycosidic groove.

It was proposed several hypotheses, revealing the role of various factors in the formation of B'-structure, but does not explain all the experimental facts. One of them [1], considering double layer "water spine" [2] of ordered water molecules in the glycosidic groove of the DNA A-tract as the principal factor stabilizing B'-conformation, is widely used by the authors of a number of works. However, this approach does not include consideration of the interaction in nonglycosidic groove, while according to experimental data the ability of the DNA segment to retain B'-structure depends on the nature of the exocyclic groups in the sixth position of purines. Moreover, volumetric effect accompanying in solution the netropsin binding with poly (dA): poly (dT) is much higher than could be explained by simple displacement of water molecules forming the "water spine" in glycosidic groove the DNA. Thus, we need for a more thorough study of hydration features of the A-tract of DNA.

# Background

The role of the water environment in the formation of double-helical DNA structure is successfully investigated by various experimental and theoretical methods for many years [3]. Studies of DNA melting and conformational transitions can be already considered classics , their results have already been included in the textbooks. However, studies of specific hydration (i.e. interactions between the atoms of DNA and water, accompanied by the formation of hydrogen bonds) held less intensively. So, back in the early 60's, localization of water molecules associated to DNA was qualitatively established using X-ray diffraction [4] and infrared spectroscopy [5] methods. In 1967, it has been hypothesized about the important role of water bridges in stabilizing the DNA double helix [6]. Since the mid-70s began an intensive study of the binding sites of the water molecules with the nucleic acid bases and the DNA fragments via computer simulation.

Studies of the structural aspects of the hydration received significant impetus to the mid-80s, after the publication in 1981 the crystalline structure of B-DNA dodecamer [7]. At this time, there are the first realistic model of water structure in the glycosidic groove of DNA [8], based both on experimental data and the results of the computer simulation. Recently various groups of scientists have made several attempts to systematically study the dependence of the structure of the water shell on the DNA nucleotide sequence and conformation. Among them are the X-ray database analysis of crystal hydrates of the DNA [9,10] and Monte Carlo simulation [11], that allowed to reveal some of the qualitative patterns in the distribution of the maxima of the local density of water around the DNA atoms.

Of course, the statistical basis of these studies is small, and the language of the structure description is ambiguous, which leads to contradictory conclusions (comparative analysis [12]). Moreover, the possibility to spread the results found for crystals is itself unclear for the DNA structure in solution. However, obtaining data on the detailed structure of the water shell of DNA becomes an urgent task, because water molecules involved in its formation can stabilize mismatched nucleotide pair [13,14], compete with various ligands [15] or, conversely, be a link in the DNA-protein recognition [16]. Our study will help to fill gaps in knowledge about the structure of the water shell and its role in stabilizing the conformational features of the double helix of B-DNA.

## Methods and algorithms for the computer experiments

To obtain quantitative estimates of the influence of hydration on the structure of the DNA double helix it is necessary to calculate of various energetical characteristics of model systems containing double-helical DNA fragment and its water shell for a representative set of the conformations of the fragment. This approach also allows to determine the power of hydration effect, depending on the nucleotide sequence. Additionally, for each system the 3D-map of distribution of water oxygen atoms around the DNA fragment, and the probability of formation of water bridges of one, two or three water molecules "connected" together and with DNA atoms by hydrogen bonds are calculated. This information will help us to reveal the location of the particular DNA atomic groups responsible for the interaction with drug molecules and enzymes.

Simulation of the water shell of the DNA fragment was carried out by Monte Carlo method (Metropolis sampling) for NVT-ensemble at room temperature. Simulated systems contain a DNA fragment consisting of 10 bp and 2,000 molecules of water (see. Fig. 1). To eliminate the surface effects the periodic boundary conditions were imposed on the system along the axis of the DNA. Intermolecular interaction energy was calculated using the atom-atom potential functions [17]. All the programs necessary for the computer simulation were written in algorithmic language FORTRAN77 using standard MPI-libraries. This allowed us to parallelize computations by two interacting processes and exploit of two processors for the computation. In the first computation process subsystem water-water was served, and in the second – the water-DNA ones.

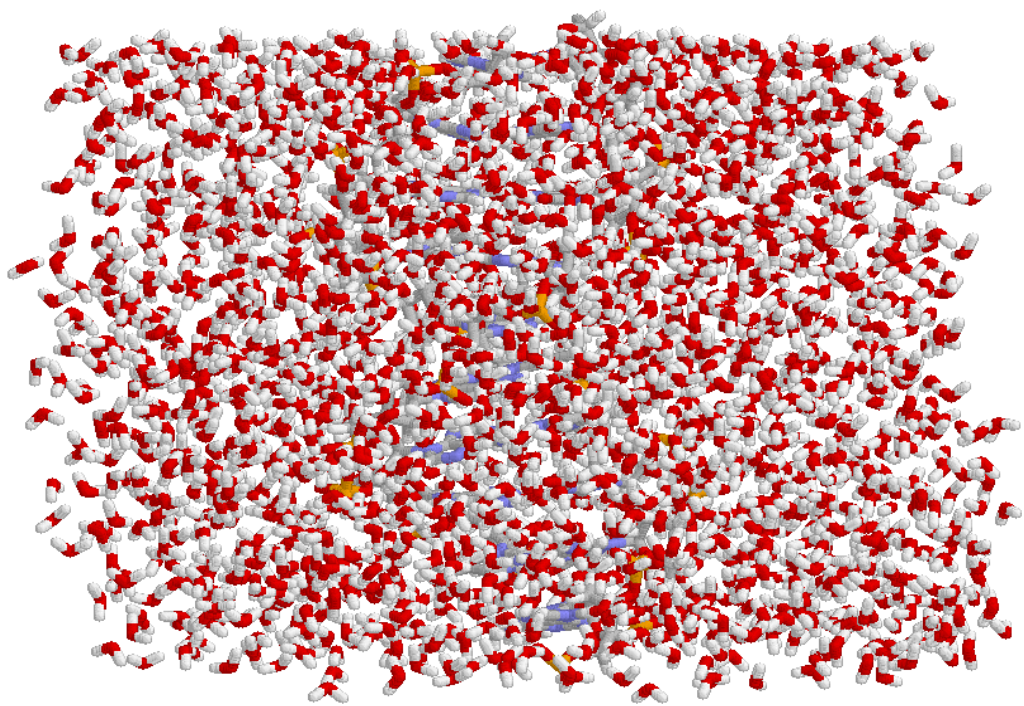

Fig. 1. A DNA fragment (blue and orange spots) surrounded by 2000 water molecules (red and white "corners"). The axis of the DNA double helix is vertical. A periodic boundary conditions are imposed on the system along the axis of the DNA helix.

We have developed a package XDNA, based on the library MPI, which allows us to calculate the energy of hydration of DNA fragments of different nucleotide sequences, as well as mapping of the spatial distribution of water molecules around a given DNA fragment. It is designed to study the regularities of the DNA structure formation in aqueous solution and can be used on computer clusters or supercomputers with parallel architecture. XDNA package allows to detect the most optimal (for the energy of hydration) structure of DNA fragments, as well as to determine the location of the DNA-binding sites preferred for water molecules and/or various ligands (e.g. drugs).

## Results of simulations

To identify patterns in the structure of the water shell of DNA containing A-tracts, we carried out a great number of computational experiments on multi-processor systems (computer cluster at IMPB RAS as well as MVS1000M at JSCC RAS). We calculated the various structural and energy characteristics for the hydration of five samples of double-helical DNA (the datasets of the DNA atoms provided by Prof. V.I. Poltev) with a nucleotide sequence as that of poly (dA): poly (dT), varying in the width of the grooves of the DNA double helix (see. Fig. 2).

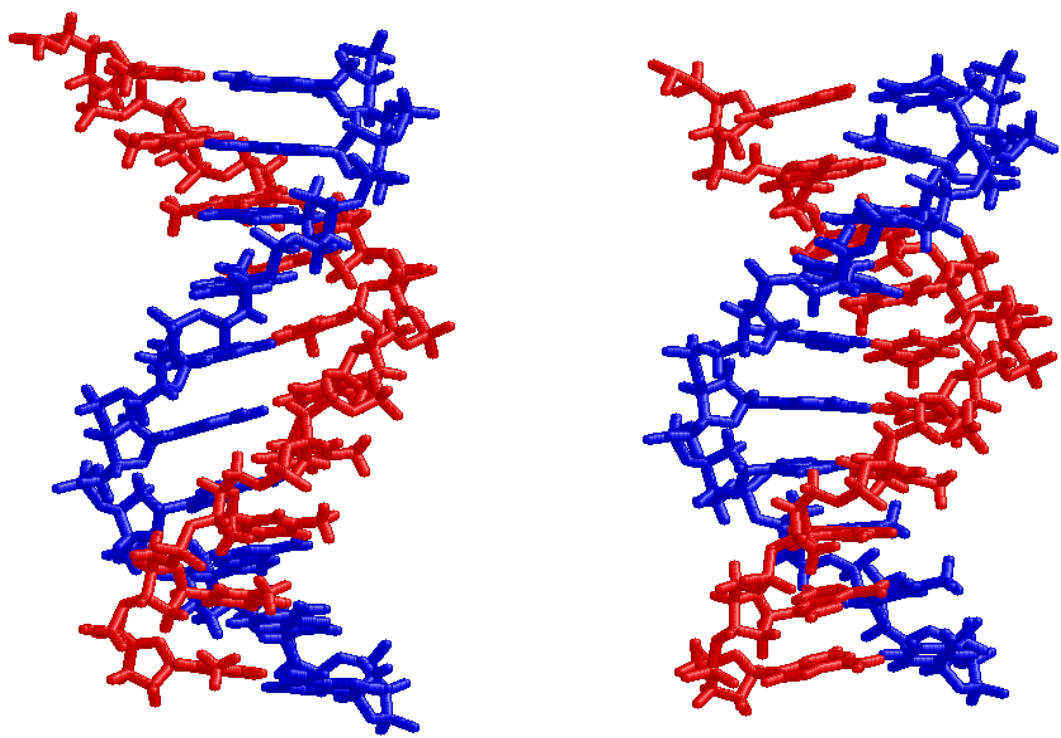

Fig. 2. B'- and B-conformations of poly (dA): poly (dT) having narrowed and widened glycosidic groove of the DNA double helix.

To obtain a reliable data in the computational experiments at least 3000000 configurations (per water molecule) were generated and for the model of B'-conformation even more – 8000000 configurations.

It is known that a number of molecules containing aromatic heterocycles (e.g., various modifications of HOECHST 33258, netropsin, distamycin, etc.) firmly bound to the double helix DNA in its glycosidic groove. This property is the basis of the mechanism of the drug effect of these substances. To date, the structures of several complexes of these molecules (ligands) with short DNA fragments were found by X-ray diffraction methods (see Protein Data Bank, structure ID's - 109D, 1QSX, 403D, 448D etc., http://www.rcsb.org/pdb/index.html). It has been established that the binding of the ligands occurs at sites of DNA containing multiple pairs A:T successively.

The developed software allows us to calculate the spatial distribution of water molecules (the probability density to detect the center of the oxygen atom of the water molecule at a given point) around the double-helical DNA fragment with a specified conformation and the nucleotide sequence (Fig. 3,4,7). Possible molecular configurations corresponding to these maps are shown in Fig. 5,6,8. According to the found in this way the spatial distribution of the oxygen atoms of water molecules the probe for the ligand's molecule can be designed specifically, increasing or decreasing the degree of compliance of their forms (and charge distributions) with the form of a predetermined site on the DNA molecular surface.

In particular, the calculated by us structure of water (Fig. 3) in the glycosidic groove of one of the possible conformations of poly (dA): poly (dT) duplex resembles a spiral ribbon of linked pentagons (Fig. 9). It also was found that the shape of ligand molecule rather accurately "fit" into the pattern (Fig. 10).

This work was supported in partly by grants of RFBR (№00-01-05000-и, №01-07-90317), INTAS 97-31753 «Design, synthesis and testing of novel biologically-active molecules as potential drugs with sequence-specific binding to nucleic acids», and No. 35239-E CONACyT (Mexica).

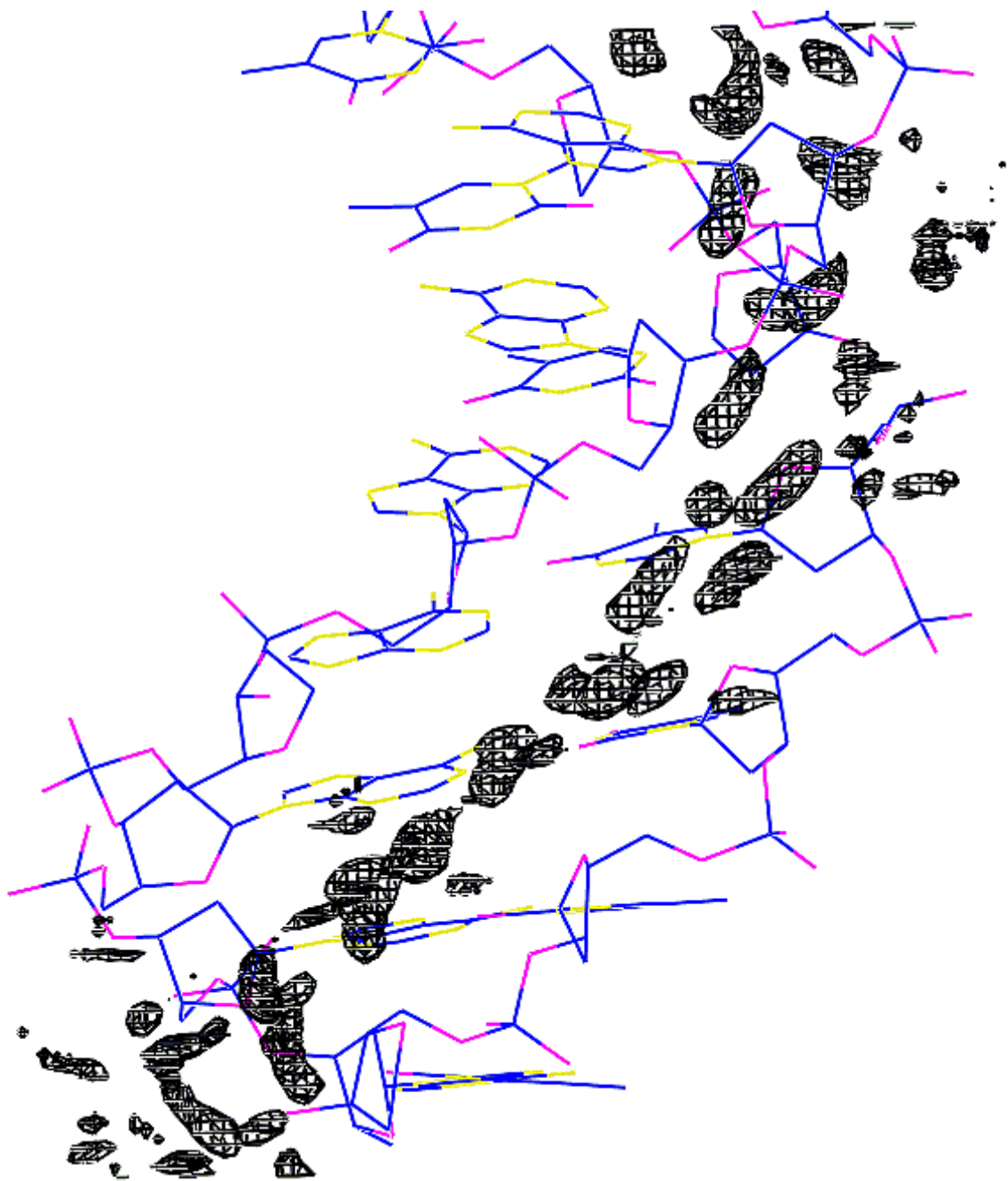

Fig. 3. The structure of water in the glycosidic groove of poly (dA): poly (dT).
B'- conformation of the DNA double helix. Width of the glycosidic groove is approximately the size of a water molecule. Black highlighted the most likely region of localization of the center of the oxygen atom of the water molecules. Top view is shown in Fig. 9 (without DNA).

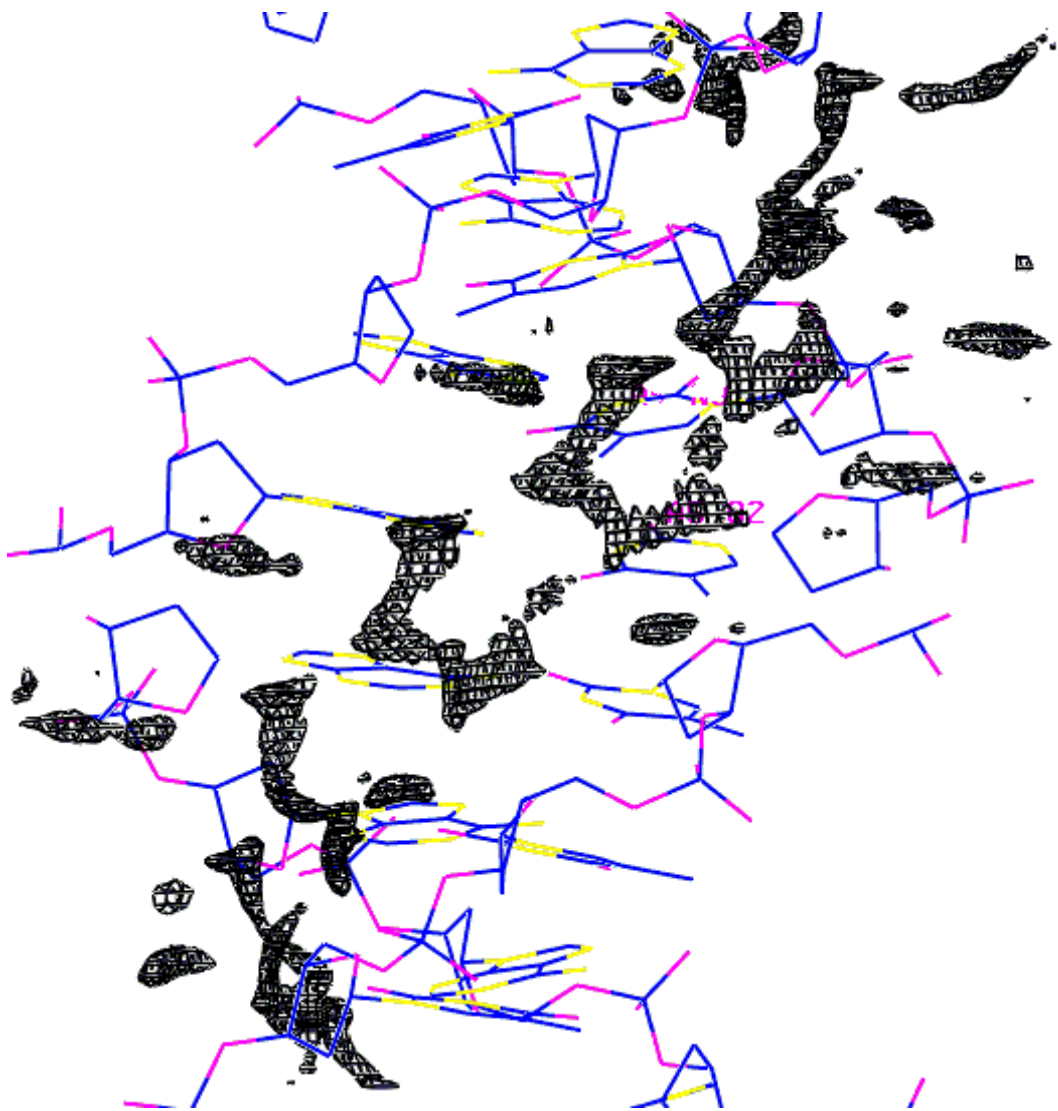

Fig. 4. The structure of water in the glycosidic groove of poly (dA): poly (dT).

B - DNA double helix conformation. The width of the glycosidic groove is slightly more than twice the size of a water molecule. Black highlighted areas of localization of centers of the oxygen atoms of water.

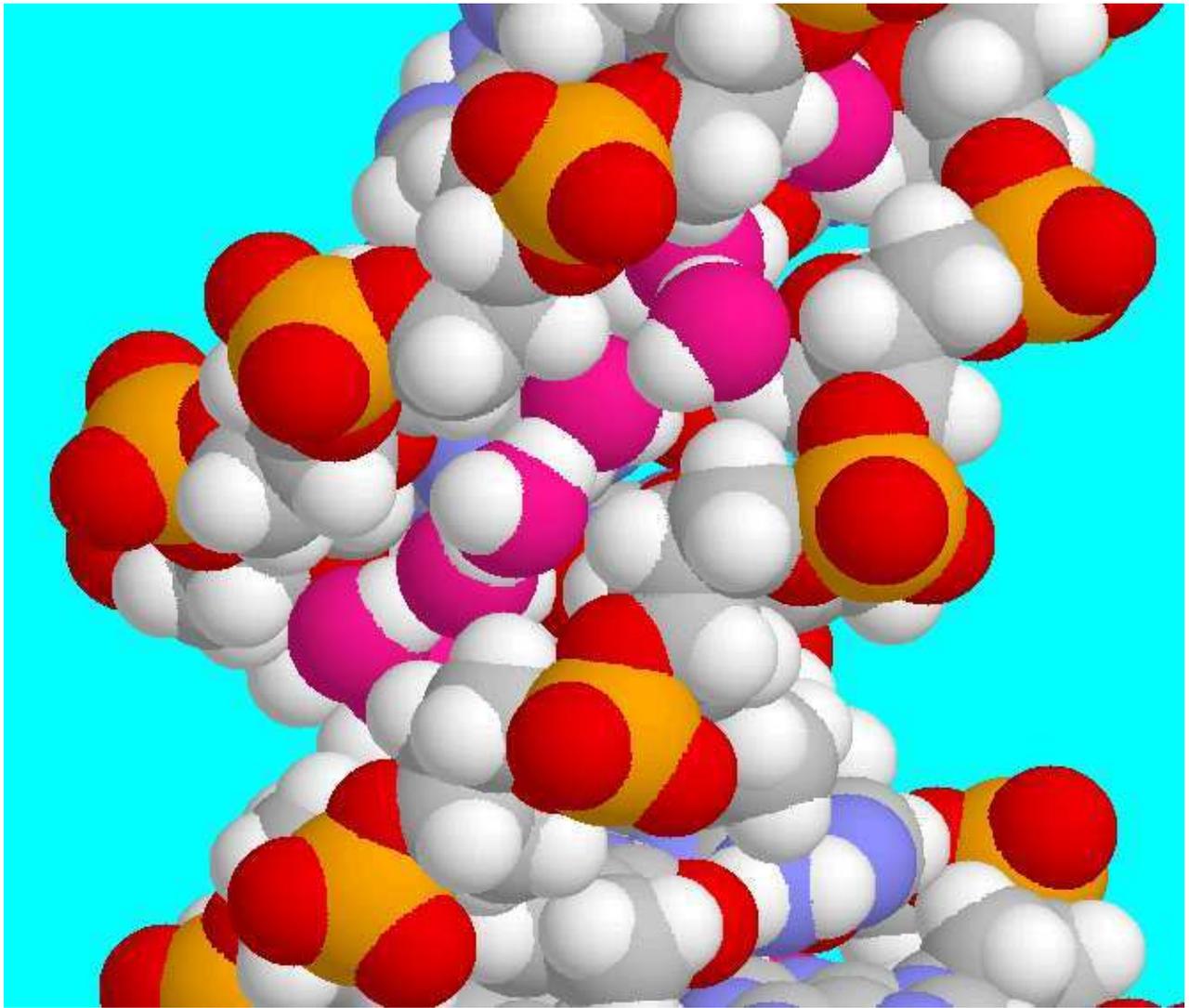

Fig. 5. View of glycosidic groove of poly (dA):poly (dT).
B'- conformation of the DNA double helix with a narrowed glycosidic groove.
The two inner layers of water molecules (magenta color) forming "water spine" are shown.

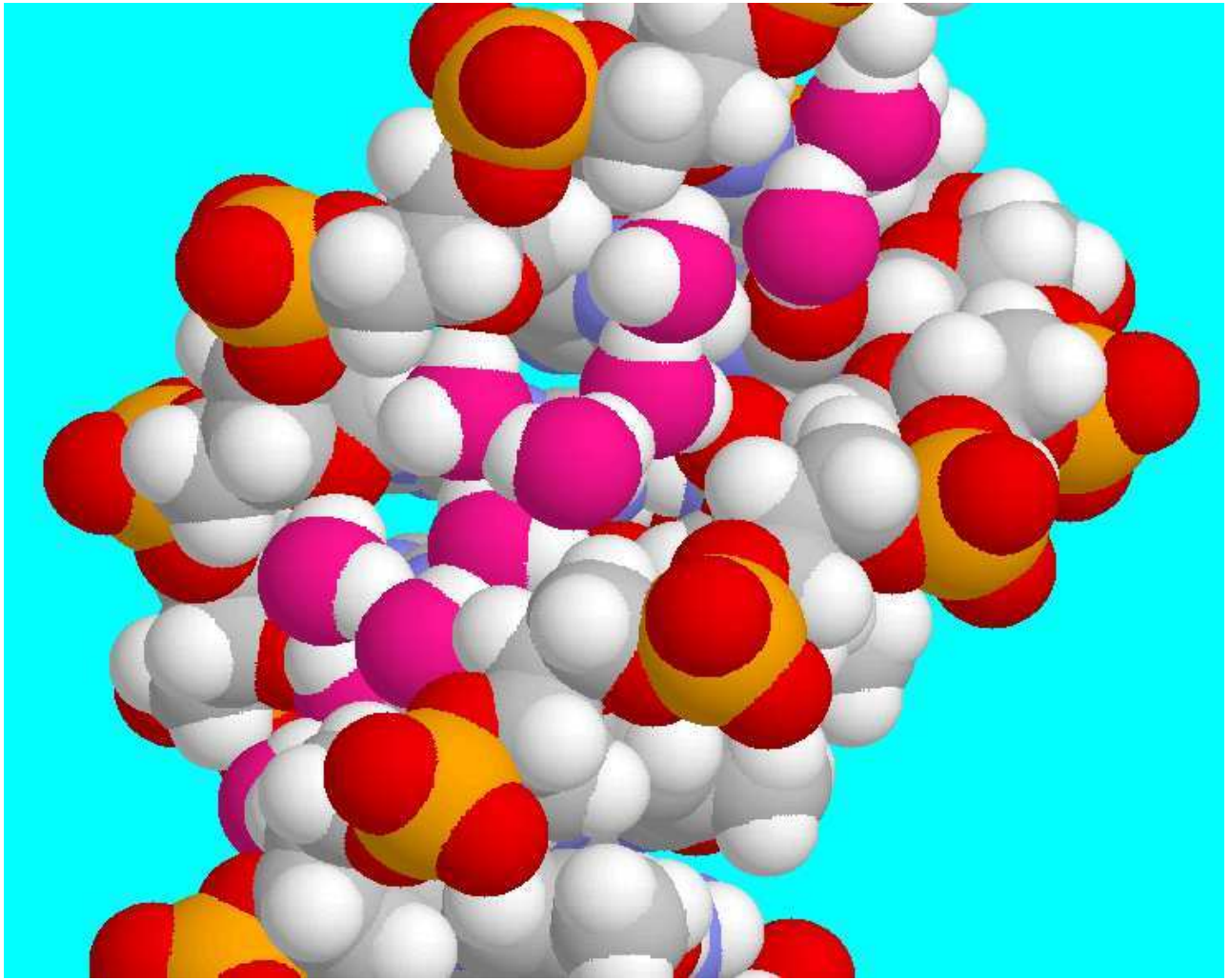

Fig. 6. View of glycosidic groove of poly (dA):poly (dT).
B- conformation of the DNA double helix with a widened glycosidic groove.
The inner hydrate layer formed of two chains of water molecules (magenta) disposed along the walls of the glycosidic groove is shown.

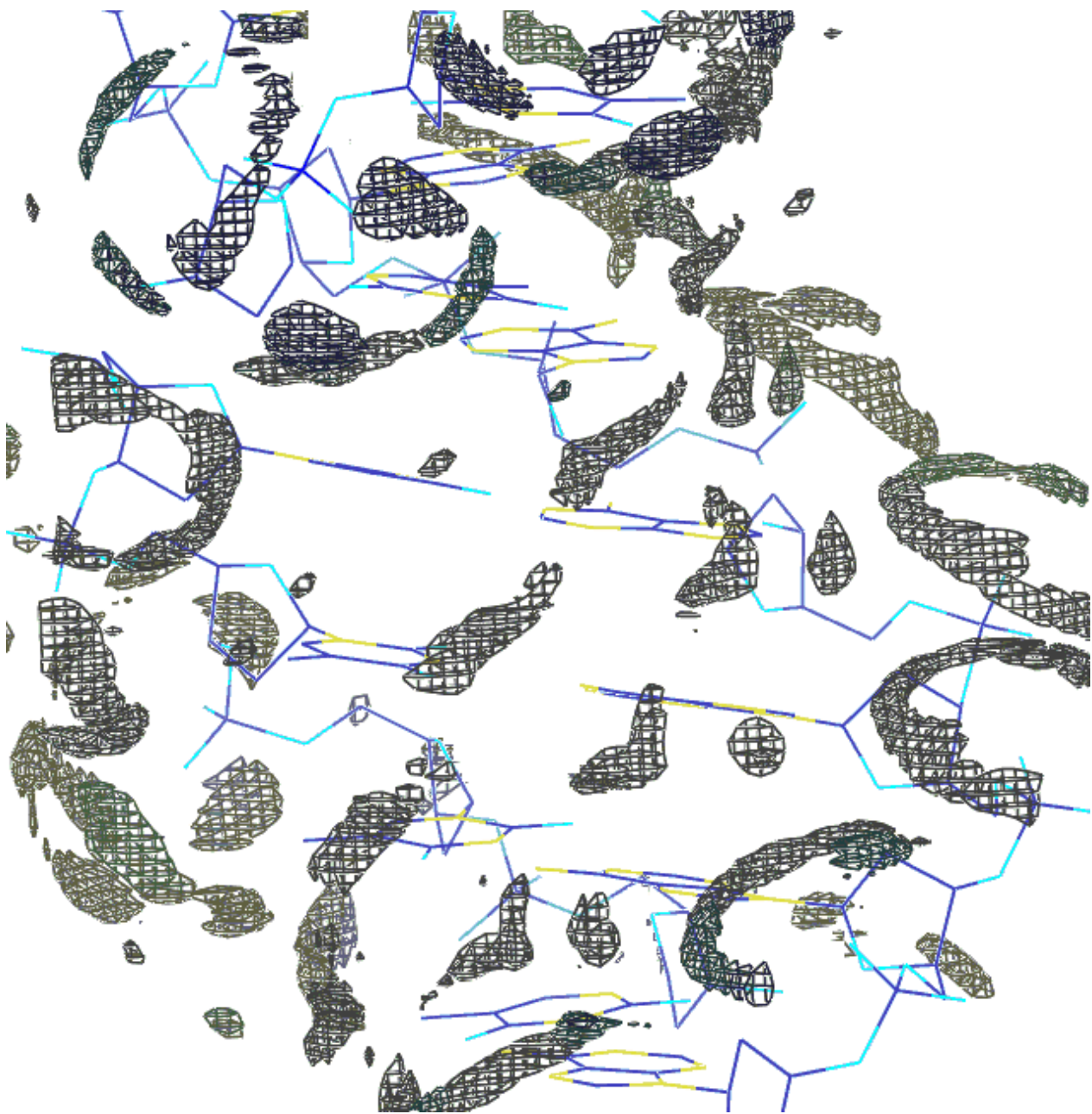

Fig. 7. The structure of water in the nonglycosidic groove of poly (dA): poly (dT). B'- conformation of the DNA double helix. Circular objects are fragments of the structure of water near the ionized phosphate groups of DNA. Black color highlighted the areas of localization of the centers of the oxygen atoms of water.

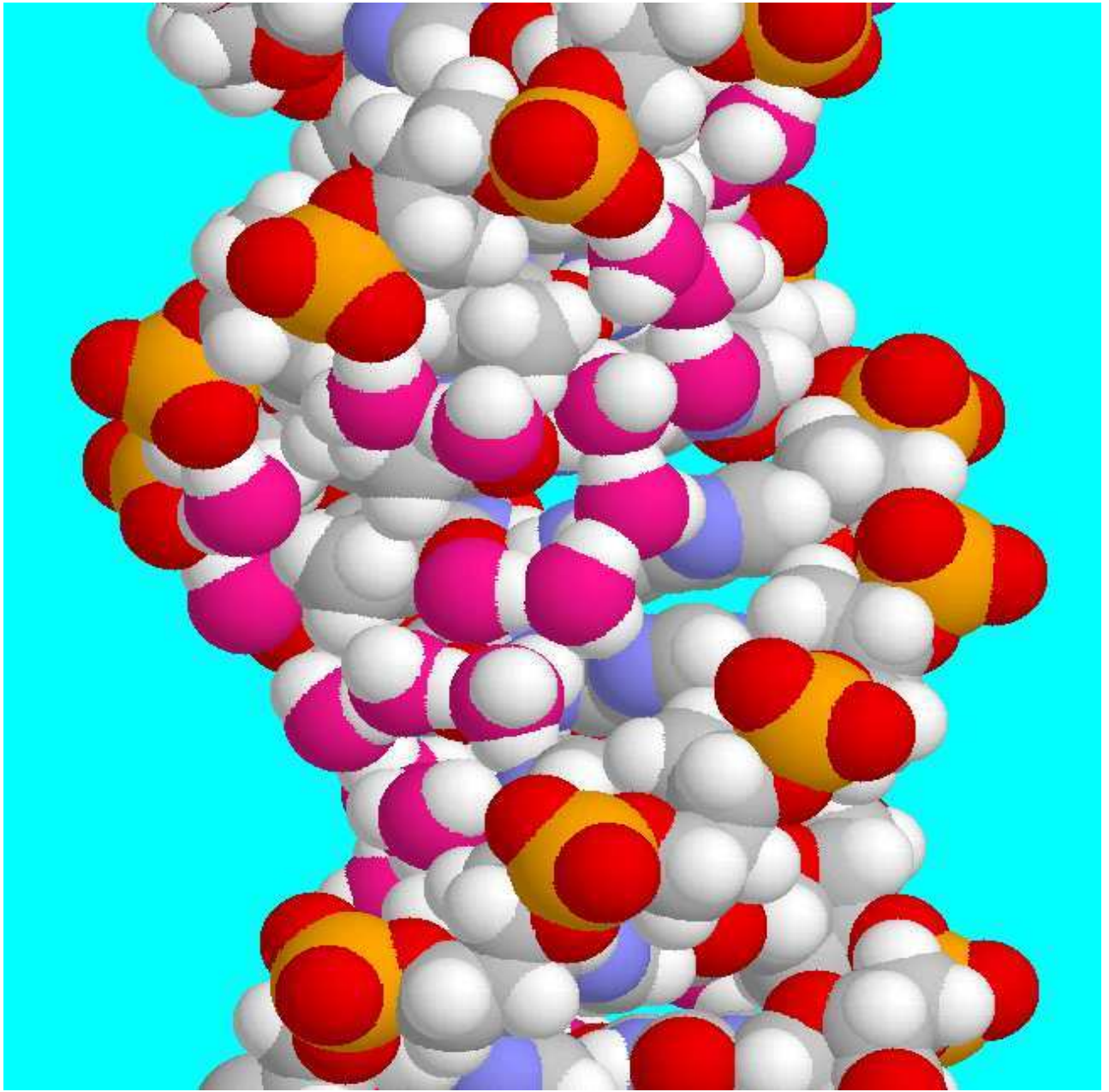

Fig. 8. View of nonglycosidic groove of poly (dA):poly (dT).
B'- conformation of the DNA double helix.
The layer of water molecules (magenta) which is closest to the "bottom" of the nonglycosidic groove of the DNA helix is shown, formed by two or three chains of water molecules located along the groove.

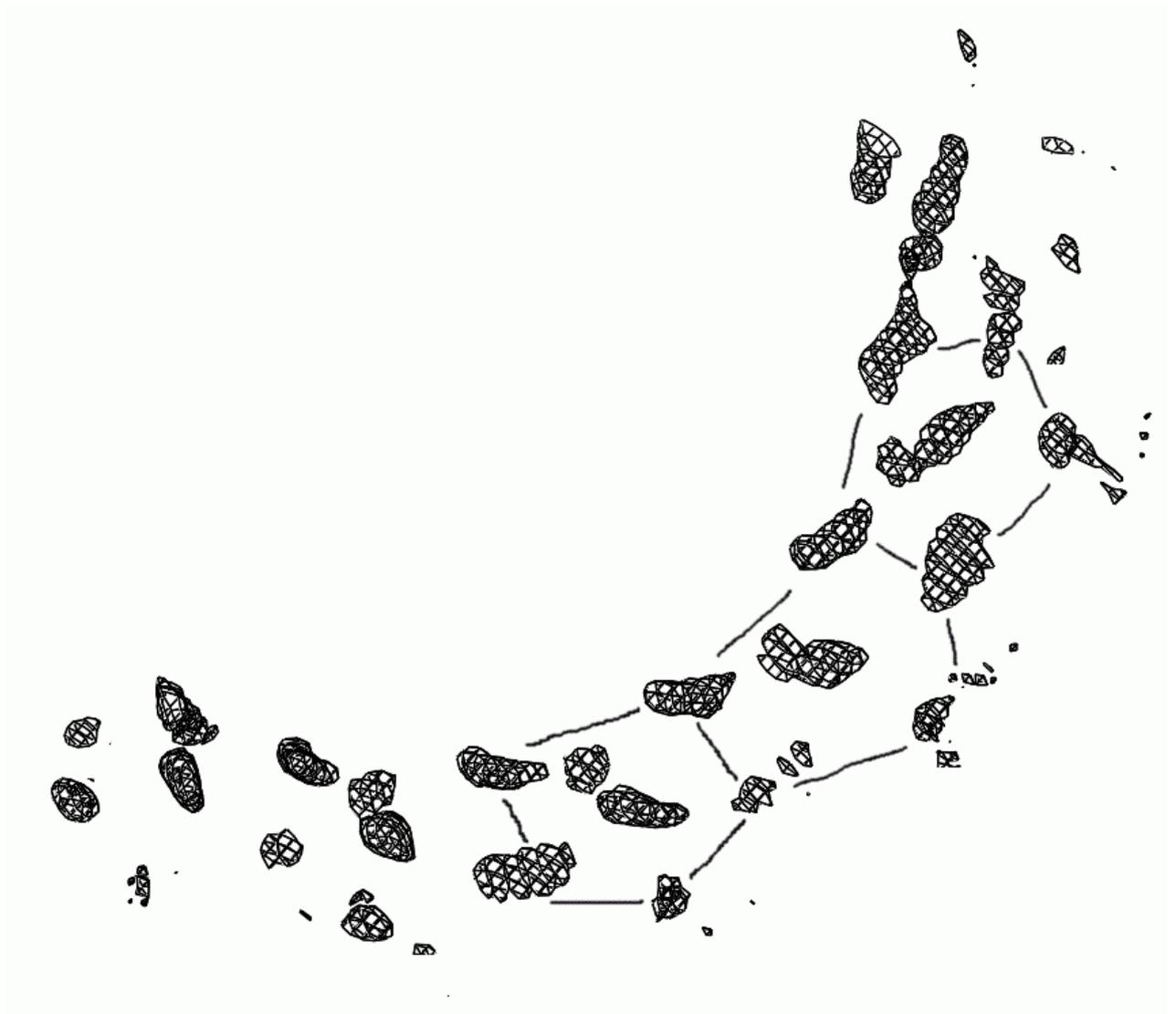

Fig. 9. Detail of the structure of water in the glycosidic groove of poly(dA):poly(dT). B'-conformation of DNA.

DNA axis is almost perpendicular to the plane of the drawing. Another projection represented in Figure 3 (with DNA). The water molecules are grouped into a flat spiral tape wrapped around the DNA double helix. Segments of "tape" are pentagons formed by water molecules (one molecule is placed in the center of the pentagon). The stability of such structures is provided by hydrogen bonds between adjacent water molecules and hydrophilic atoms of the DNA.

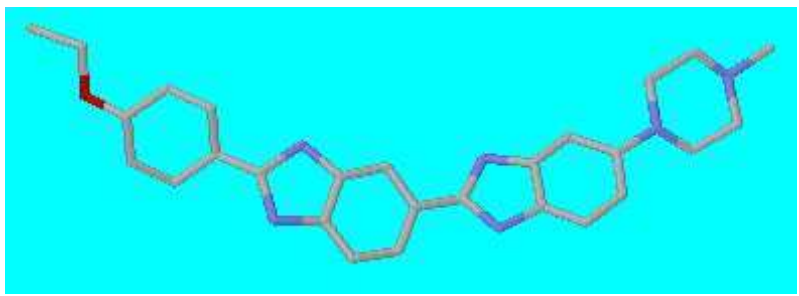

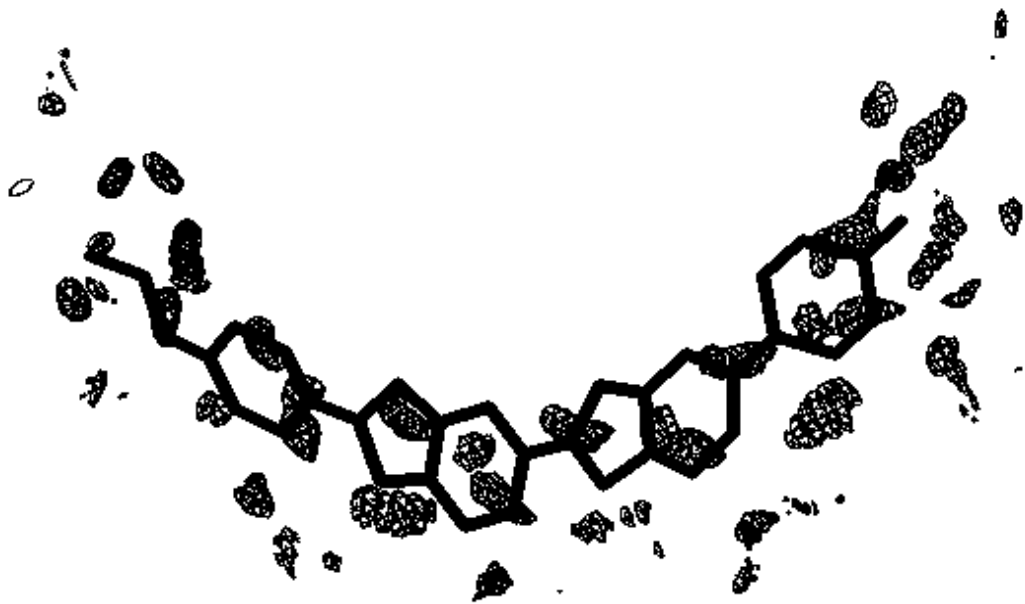

Fig. 10. Structural compliance of aqueous "tape" from the glycosidic groove of B'-conformation of poly (dA): poly (dT) and the shape of the molecule of the antibiotic of bis-benzimidazole type (the upper part of the figure, hydrogen atoms are not shown).